\documentstyle[aps,epsf]{revtex}
 \newcommand{\zb}{\mbox{$\beta$}}  
  
\newcommand{\ze}{\mbox{$\epsilon$}}

\newcommand{\zl}{\mbox{$\lambda$}}

 \newcommand{\zD}{\mbox{$\Delta$}}

\newcommand{\xa}{\alpha} \newcommand{\xb}{\beta}

\newcommand{\xl}{\lambda}

\newcommand{\scr}[1]{\mbox{$\cal #1$}}

\newcommand{\itv}[1]{\mbox{\boldmath ${#1}$}}

\newcommand{\JMP}[3] {J.\ Math.\ Phys.\ {\bf{#1}}, {#2}  ({#3})}

\newcommand{\JPC}[3] {J.\ Phys.\ C{\bf{#1}}, {#2} ({#3})}

\newcommand{\JPSJ}[3] {J.\ Phys.\ Soc.\ Jpn.\ {\bf{#1}}, {#2}  ({#3})}

\newcommand{\PR}[3] {Phys.\ Rev.\ {\bf{#1}}, {#2}  ({#3})}

\newcommand{\PRB}[3] {Phys.\ Rev.\ B{\bf{#1}}, {#2}  ({#3})}
\newcommand{\PRL}[3] {Phys.\ Rev.\ Lett.\ {\bf{#1}}, {#2}  ({#3})}

\newcommand{\Physica}[3] {Physica {\bf{#1}}, {#2}  ({#3})}
\newcommand{\PTP}[3] {Prog.\ Theor.\ Phys.\ {\bf{#1}}, {#2}  ({#3})}
\newcommand{\RMP}[3] {Rev.\ Mod.\ Phys.\ {\bf{#1}}, {#2}  ({#3})}
\newcommand{\SSC}[3] {Solid State Commun.\ {\bf{#1}}, {#2}  ({#3})}
\newcommand{\ZPB}[3] {Z.\ Phys.\ B{\bf{#1}}, {#2}  ({#3})}
\newcommand{\ZP}[3] {Z.\ Phys.\ {\bf{#1}}, {#2}  ({#3})}

\begin{document}
\title
{
Dynamical Effective Medium Theory for Quantum Spins and Multipoles
} 
\author
{
Yoshio {\sc Kuramoto}\footnote{e-mail address: kuramoto@cmpt01.phys.tohoku.ac.jp} and Noboru {\sc Fukushima}
}

\medskip
\address
{
Department of Physics, Tohoku University, Sendai 980-77
}
%
%\today
\maketitle

\medskip
\begin{abstract}
A dynamical effective medium theory is presented for quantum spins and higher multipoles such as quadrupole moments.
The theory is a generalization of the spherical model approximation for the Ising model, and is accurate up to $O(1/z_n)$ where $z_n$ is the number of interacting neighbors.
The polarization function is optimized under the condition that it be diagonal in site indices.  
With use of auxiliary fields and path integrals, the theory is flexibly applied to quantum spins and higher multipoles with many interacting neighbors.
A Kondo-type screening of each spin is proposed for systems with extreme quantum fluctuations but without conduction electrons. 

\end{abstract}

%
%\clearpage

\section{Introduction}

Strongly interacting localized electrons have been attracting renewed interest.  A key feature characterizing the new development is the coupling of spin and orbital degrees of freedom.
For example, Ce$_{1-x}$La$_x$B$_6$ \cite{goto,sakakibara,sera} 
has both magnetic and electric quadrupole orders, and
the phase diagram exhibits intriguing systematics as $x$ varies.
The specific heat above the quadrupolar ordering temperature shows a large contribution of fluctuation, which is suppressed by applied magnetic field. \cite{sera}  
Similar but non-identical behavior has been found in TmTe.\cite{matsumura}  We note that Ce$_{1-x}$La$_x$B$_6$ 
shows the Kondo effect in the resistivity,  while TmTe is insulating. 
In 3d systems, the orbital degrees of freedom  appears as the static and dynamic Jahn-Teller effects. \cite{khomskii}  
An exemplary system is La$_{1-x}$Sr$_x$MnO$_3$,  which shows the drastic change of magnetic anisotropy \cite{endoh} with increasing $x$.  
The transport property is characterized by
the colossal magneto-resistance. \cite{tokura}
All these phenomena require simultaneous account of orbital and spin degrees of freedom.

These developments motivate construction of a new quantum theory which can deal with fluctuation effects of not only spin but higher-order multipoles.
Concerning previous efforts toward the dynamical theory, we refer to the work of Hubbard\cite{hubbard} which addresses the high-temperature limit of the Heisenberg model in high dimensions.  
For the low-dimensional Heisenberg model, highly sophisticated theories are available with account of quantum fluctuations.\cite{tsvelick}  However, these theories are not suitable to three-dimensional systems with a different character of fluctuations.
 
In the case of strongly correlated itinerant electrons, microscopic theories have been developed with account of both the Kondo effect and the intersite correlation for the Anderson lattice, \cite{kuramoto85,kuramoto87,kim,georges}  and for the Hubbard model. \cite{georges,vollhardt,erwin}
These theories use the idea of dynamical effective medium which is justified in the limit of large spatial dimensions.
Although the approach is highly successful in deriving the density of states of
 electrons with nontrivial structure around the Fermi level, 
the magnetic property is treated rather crudely. 
To clarify the need of improvement, let us take an example of the half-filled Hubbard model.
With large Coulomb repulsion the single-particle spectrum has an energy gap.
As a result the spectrum of the effective medium to determine the Green function also has a gap.  
Under this condition the infinite-dimensional theory predicts a gapful spectrum also for spin excitations.  In reality, however, the spin excitation is gapless in many cases even though the single-particle spectrum has a gap.
Furthermore in the insulating paramagnetic phase the entropy derived by the theory does not vanish at zero temperature.  
With a magnetic ordering, the entropy can vanish by the same mechanism as the band antiferromagnetism.  However, in the limit of vanishing charge fluctuation,  one has complete spin polarization in the ground state.  This last feature of the theory is not satisfying since quantum spin fluctuations should reduce more or less the magnitude of ordered moments.
In order to remedy the situation one has to go beyond the lowest-order self-consistent theory.  Then there appears a difficulty of analyticity in the Hubbard model and other fermion models. \cite{ingersent}

The purpose of this paper is to present a next-leading order self-consistent theory which is free from the above deficiency.  The formulation has so far been attained only for systems with localized electrons such as quantum spins. 
The present theory can deal with a possible paramagnetic ground state without residual entropy, provided that the spin excitation has a gapless spectrum.
The paper is organized as follows: In the next section we take the Ising model as the simplest system to apply the formalism.  We exploit the variational character of the thermodynamic potential in deriving the magnetization and the susceptibility.
The same variational property is also used to compute the entropy and specific heat within the same scheme.
Section 3 describes the extension of the formalism to quantum models.  We take the Heisenberg model with arbitrary exchange interactions as a representative quantum model.  The static approximation is introduced which is much simpler to treat than the fully dynamical counterpart.  The approximation is justified in the high temperature limit and makes a correspondence to the classical theory.
In \S 4 we discuss a way to solve the problem by mapping to 
the spin-boson system where an impurity spin interacts with bosonic environment.  The latter represents the dynamical medium provided by surrounding spins.  
By this mapping, numerical methods such as the Monte Carlo technique, numerical renormalization group,  or the self-consistent perturbation theory become applicable to solve
the impurity problem.
Section 5 applies the theory to more general systems with quadrupole moments or with crystalline-electric-field effects.
In the final section we discuss the results, in particular a possibility of a Kondo-type effect in spin systems without conduction electrons.
 
\section{Self-Consistent Renormalization for the Ising Model}

\subsection{Perturbation theory with auxiliary fields}

Although our principal interest is in quantum models, we first take the Ising model to present the new formalism in the simplest, and hence the most transparent manner.  
The spherical model approximation (SMA) goes one step beyond the mean-field theory for the Ising model, and has been discussed for many years. \cite{mattis}
We rederive the SMA in a variational formalism, which makes clear the structure as well as limitation of the approximation.
As a byproduct, we derive a formula to express the specific heat in terms of the susceptibility.

The Hamiltonian of the Ising model is given by
\begin{equation}
H = -\frac 12\sum_{ij}J_{ij}\sigma_i \sigma_j
-\sum_{i}h_i\sigma_i ,
\end{equation}
where each $\sigma_i$ takes $\pm 1$ and $h_i$ denotes a magnetic field at site $i$.
For clarity we assume that the interactions $J_{ij}$ are all positive, and the inverse of the matrix $\{J_{ij}\}$ exists.
The self-consistency equations, however, are valid for more general interactions.
We introduce auxiliary fields
which obey the Gaussian distribution, and which couple with $\sigma_i$ locally. 
The basic identity is written as
\begin{equation}
\exp\left( \frac \beta 2\sum_{ij}J_{ij}\sigma_i \sigma_j\right) = \left(\det \beta J\right)^{-1/2}
\prod_l\int _C
\frac{\beta d\phi_l}{\sqrt{2\pi}} \exp\left[- \frac \beta 2\sum_{ij}(J^{-1})_{ij}\phi_i \phi_j +\beta \sum_i \phi_i \sigma_i \right],
\label{eq:Gaussian}
\end{equation}
where the the integration path $C$ runs from $-R\exp (i\pi/4)$ to $R\exp (i\pi/4)$ with $R$ going to infinity.
This choice of a path makes it possible to obtain the convergent integral after the change of integration variables.  Namely we distort the path either along the whole real axis or the imaginary axis depending on the sign of the eigenvalue of the matrix $\{ J_{ij}\}$. 
The partition function is written as
\begin{equation}
Z  =  \sum_\sigma\int \scr D \phi \exp \left(-\beta H_\phi\right),
\end{equation}
where
\begin{equation}
H _\phi  = \frac 12\sum_{ij} (J^{-1})_{ij}\phi_i\phi_j
-\sum_i (\phi_i +h_i)\sigma_i .
\label{eq:phi-term}
\end{equation}
Thus we see that the Gaussian identity casts the original two-body interaction into the sum of local coupling terms.

In order to perform perturbation theory we first set
the energy scale to $\sum_j J_{ij}$ which corresponds to the ferromagnetic transition temperature in the mean-field approximation (MFA).
Thus each $J_{ij}$ is of order $1/z_n$ where $z_n$ is the number of equivalent neighbors.  
To organize the perturbation series concisely, it is convenient to use the fermion representation for Ising spins.  Namely we introduce spinless fermions by
\begin{equation}
\sigma _i = 2f_i^\dagger f_i -1 ,
\end{equation}
where $f_i$ is the annihilation operator of a fermion at site $i$.
Then each term of the perturbation expansion can be expressed in terms of Feynman diagrams.  
The intersite contribution involves the bare cumulant $\langle \phi_i\phi_j\rangle_0 =TJ_{ij}$.
We remark that the same diagrams appear when we use the fermion representation from the outset without introducing the $\phi$ field.  However for quantum models the $\phi$ field makes the theory much more concise.

Figure 1(a) shows the zero-th order contribution to the polarization function which corresponds to the self-energy of  the renormalized exchange interaction.   
The small parameter $1/z_n$ is offset by the site summation in these diagrams which are called tadpoles.  The tadpoles lead to shift of the effective fermion level, and represent as such the molecular field.
%%%%%%%%%%%%%%Figure%%%%%%%%%%%%%%%%%%%%%%
\begin{figure}
\epsfxsize = 15cm \epsfbox{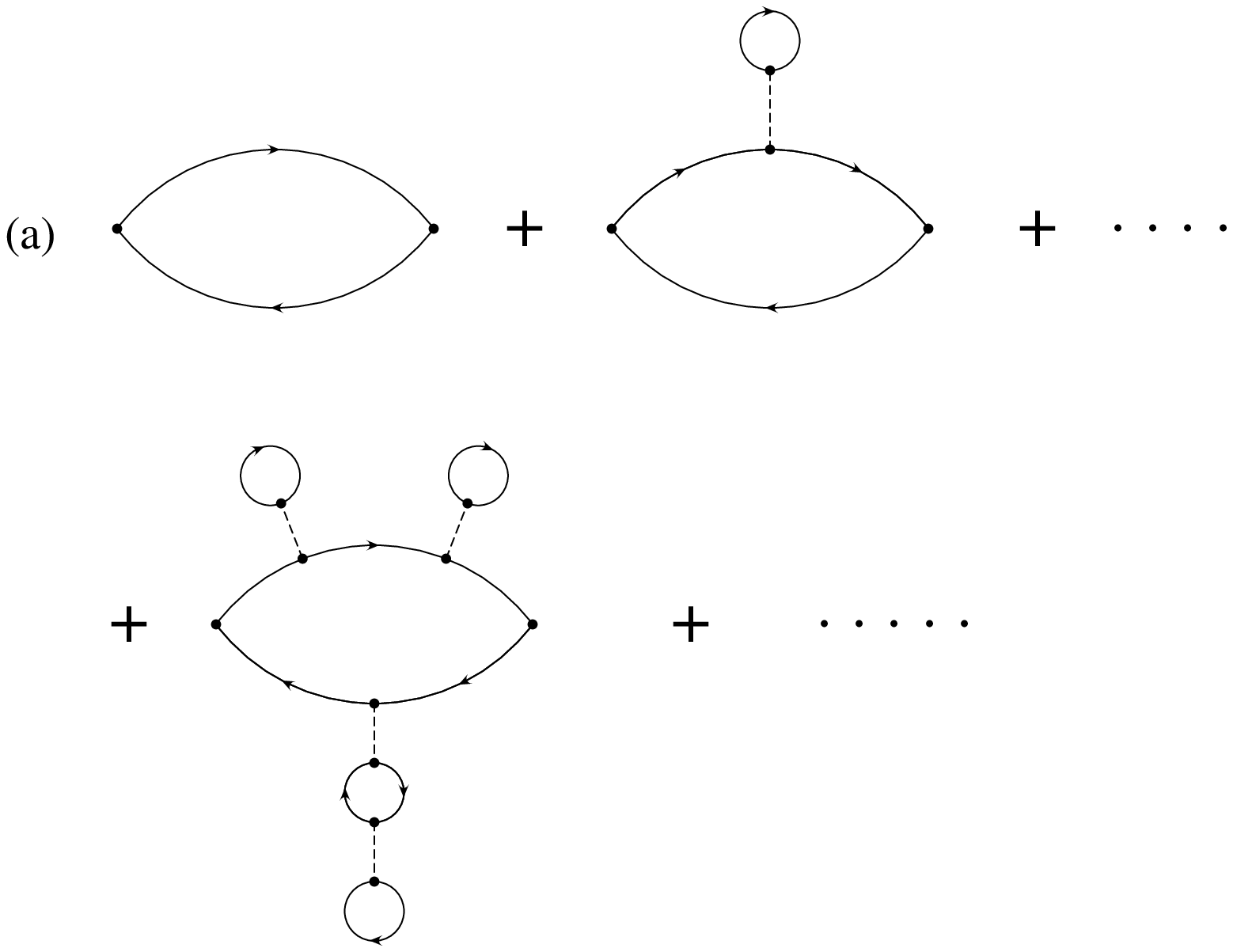}
\vspace{5mm}
\epsfxsize = 15cm \epsfbox{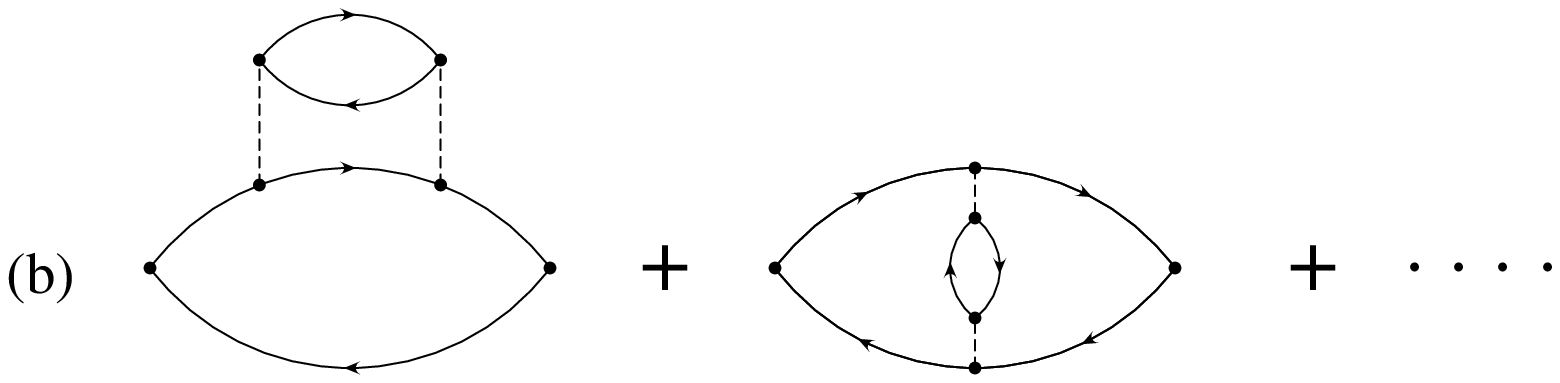}
\vspace{5mm}
\epsfxsize = 5 cm \epsfbox{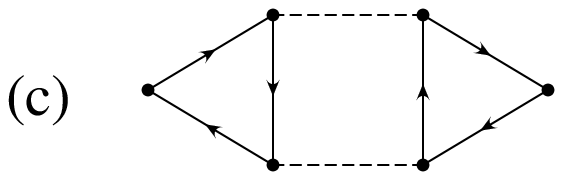}
\vspace{0.5cm}
\caption{ Feynman diagrams for the polarization function. (a) Contributions of the zeroth order in $1/z_n$; (b) first order; (c) second order.  The solid lines represent the Green function of fermions, and the dashed lines do the exchange interaction.
}
\label{fig.1}
\end{figure}
 %%%%%%%%%%%%%%%%%%%%%%%%%%%%%%%Figure%%%%%%%%%%%%%%%%%%%%%%
The set of zero-th order diagrams make a tree-like structure composed of tadpoles.  In the leading order there are no loops composed of interaction lines.
Summation of all the tree diagrams is equivalent to the saddle point approximation of eq.(\ref{eq:phi-term}). 
Namely after taking the trace over $\sigma_i$ in eq.(\ref{eq:phi-term}), we find that the saddle point $a_i$ of the field $\phi_i$ satisfies the condition
\begin{equation}
\sum_j (J^{-1})_{ij}a_j -\tanh \beta (h_i+a_i ) =0.
\end{equation}
It is seen that $a_i$ gives the molecular field at site $i$, and the magnetization $m_i =\langle\sigma_i\rangle $ is given by $m_i= \tanh \beta (h_i+a_i)$.
 Namely we get
\begin{equation}
a_i = \sum_j J_{ij} m_j . 
\end{equation}

Correction to the saddle point approximation is higher order in the small parameter $1/z_n$.
Examples of diagrams of $O(1/z_n)$ is shown in Fig.1(b).   In addition to the self-energy correction to the fermion level, vertex correction appears which contributes to the local field beyond the MFA.  Note that to this order the polarization function is diagonal in site indices.  Only from the order of $1/z_n^2$,  there emerge off-diagonal contributions.  An example is shown in Fig.1(c).  
Thus we see that in the limit of large dimensions the MFA becomes exact, and that the polarization function $\Pi _i$ is diagonal in site indices to the next leading order.
The renormalized exchange interaction $\bar{J}_{ij}$ is given by
\begin{equation}
\left(\bar{J} ^{-1}\right)_{ij} = \left(J^{-1}\right) _{ij} -\Pi _i\delta_{ij}.
\label{eq:chi_feyn}
\end{equation}
Once the renormalized exchange is determined, the 
 (differential) magnetic susceptibility is derived as the two-particle Green function of the fictitious fermions. Namely the (renormalized) cumulant average is related to the susceptibility as 
\begin{equation}
\chi_{ij} = 4\beta \langle f_i^\dagger f_i f_j^\dagger f_j \rangle _c
 = \beta \langle \delta\sigma_i \delta\sigma_j\rangle, 
\end{equation}
where $\langle\cdots\rangle_c$ means the cumulant average, and
$\delta\sigma_i = \sigma_i -\langle\sigma_i \rangle .$ 
In terms of the polarization function the susceptibility satisfies the Dyson-type equation:
\begin{equation}
\chi _{ij} = \Pi _i(\delta_{ij}  +\sum_l J_{il}\chi _{lj}).
\label{eq:dyson}
\end{equation}
This equation is also written in the matrix form as
\begin{equation}
\chi ^{-1} =\Pi ^{-1} -J.
\end{equation}
In the next subsection we derive the susceptibility explicitly.

\subsection{optimization of polarization function}

In a theory accurate only up to $O(1/z_n)$, there is some ambiguity how to include higher-order terms.  
We respect the self-consistency or, equivalently, the variational property of the theory in choosing higher-order terms. \cite{kuramoto87}
Let us consider the perturbation expansion of the thermodynamic potential $\Omega = -T\ln Z$. 
In order to make a renormalized expansion, 
we  introduce a fictitious field $u_{ij}$ which shifts $(J^{-1})_{ij}$ to $ (J^{-1})_{ij} +u_{ij}$ in $H_\phi$.
With infinitesimal change of $u_{ij}$,  the corresponding change of $\Omega$ occurs as 
\begin{equation}
2\delta \Omega = \sum_{ij}\langle \phi _i \phi _j \rangle _c\delta u_{ij}. 
\label{eq:delta Omega_c}
\end{equation}
Here we consider the case where $\delta u_{ij}$ couples on the average with only the connected part of  $\langle \phi _i \phi _j \rangle$, {\it i.e.} the cumulant average:
$$
\langle \phi _i\phi _j  \rangle _c = 
\langle \phi _i\phi _j  \rangle  - 
\langle\phi _i \rangle \langle \phi _j \rangle =T\bar{J}_{ij}.
$$ 
An example of external fields without coupling to the disconnected part $\langle\phi _i \rangle \langle \phi _j \rangle$ 
 is the case where $\delta u_{ij}$ has a modulation corresponding to a finite momentum with $h_i$ constant.

The Dyson equation gives the relationship between $\bar{J}_{ij}$ and the bare exchange as
$\bar{J}^{-1}  = J^{-1} +u-\Pi$ 
in the matrix notation.
By using 
$\delta u = \delta\bar{J}^{-1} +\delta \Pi$,  we can eliminate the fictitious field.   Then we get
\begin{equation}
2\beta\delta \Omega = {\rm Tr}\delta  (\ln \bar{J}^{-1}+\bar{J}\Pi )-\delta \Phi \{\bar{J}\} ,
\label{eq:delta Omega}
\end{equation}
where Tr is the trace over the site index and the functional
$\Phi\{\bar{J}\}$ is introduced so as to give 
$ \delta \Phi/\delta \bar{J}  = \Pi $.   
Equivalently,  $\Phi \{\bar{J}\}$ is written symbolically as
\begin{equation}
\Phi \{\bar{J}\} = {\rm Tr} \sum_{n=0}^{\infty} \frac{1}{n+1} \Pi_n \{\bar{J}\} \bar{J},
\label{eq:skeleton}
\end{equation}
where $\Pi_n $ is the polarization function made up of all $n$-th order skeleton diagrams with respect to $\bar{J}$.
Upon variation of $\bar{J}$ in an $n$-th order skeleton,  we have $n$ ways of choosing $\bar{J}$.  Hence the denominator is cancelled in the final result, and $\delta\Phi/\delta\bar{J}$ in eq.(\ref{eq:skeleton}) indeed gives $\Pi$.
Now we integrate eq.(\ref{eq:delta Omega}) with respect to the exchange interaction ranging from $J$ to $\bar{J}$.  
The result is given by 
\begin{equation}
2\beta(\Omega -\Omega_0) =  {\rm Tr}\left[ \ln (\bar{J}^{-1}J)  +\bar{J}J^{-1} -1\right]  
-\Phi \{\bar{J}\} ,
\label{eq:Omega}
\end{equation}
where 
$$
\Omega_0 = \frac 1 2\sum_{ij}(J^{-1})_{ij}a_i a_j -T\sum_i\ln [2\cosh \beta (h_i +a_i) ],
$$ 
comes from the lower limit of integration and corresponds to the MFA.
We interpret eq.(\ref{eq:Omega}) as giving a variational form which is minimized by the optimum $\bar{J}$.
Since $\Pi$ is diagonal in site indices, only the diagonal part of $\bar{J}_{ij}$ enters the expansion eq.(\ref{eq:skeleton}).
Furthermore with the optimum choice $\bar{J}^{-1} = J^{-1}-\Pi$, 
we have ${\rm Tr}(\bar{J}J^{-1} -1) = {\rm Tr}(\bar{J}_D\Pi)$ where $\bar{J}_D$ is the diagonal part of the matrix $\bar{J}$. 
Thus except for the term $\ln (\bar{J}^{-1}J)$, only the diagonal part $\bar{J}_D$ of $\bar{J}$ is relevant to $\Omega-\Omega_0$.

The foregoing rearrangement guides us how to select terms of higher order for self-consistency of the theory.
For explicit calculation of $\Omega$ we decompose $H_\phi$ as
\begin{equation}
H_\phi = H_G +\sum_i H_i -H_D, 
\end{equation}
where
\begin{eqnarray}
H _G & = & \frac 12\sum_{ij} (\bar{J}^{-1})_{ij}\delta\phi_i\delta\phi_j
+\frac 12\sum_{ij}(J^{-1})_{ij}a_i a_j ,\\
H_i & = & \frac 12 \tilde{J}^{-1}_i\delta\phi_i^2 
+(m_i-\sigma_i)\delta\phi_i -(h_i+a_i)\sigma_i, \\
H _D & = & \frac 12\sum_i (\bar{J}_{ii})^{-1}(\delta\phi_i)^2.
\end{eqnarray}
Here $ \delta\phi _i = \phi _i -a_i $
is the deviation from the average value $a_i$, 
$\tilde{J}_i^{-1} = (\bar{J}_{ii})^{-1}+\Pi_i$ and $m_i =\sum_j (J^{-1})_{ij} a_j $.
The quantities $\bar{J}_{ij}$ and $a_i$ are to be determined variationally.
We regard $H_G$ as the unperturbed Hamiltonian.
Namely we have
\begin{equation}
Z = Z_G\langle \exp [-\beta (\sum_i H_i -H_D)]\rangle _G ,
\label{eq:Z product}
\end{equation}
where $Z_G ={\rm Tr}\exp (-\beta H_G)$ and
the average is taken with respect to the distribution 
$\exp (-\beta H_G)$.
The unperturbed thermodynamic potential $\Omega_G =-T\ln Z_G$
is given by
\begin{equation}
\Omega_G  = -\frac T2 {\rm Tr} \ln (\bar{J}J^{-1}) +\frac 12\sum_{ij}(J^{-1})_{ij}a_i a_j ,
\label{eq:Omega_G}
\end{equation}
where $\ln J^{-1}$ enters via the normalization given by eq.(\ref{eq:Gaussian}).
By comparing eqs.(\ref{eq:Omega}), (\ref{eq:Z product}) and (\ref{eq:Omega_G}), 
we see that the average in eq.(\ref{eq:Z product}) should rather be taken with use of $\exp(-\beta H_D)$ in order to achieve self-consistency up to $O(1/z_n)$. 
Thus we obtain the partition function as
$
Z = Z_G Z_L/Z_D.
$ 
Here $Z_L$ is the product of local contributions given by
\begin{equation}
Z _L =  \prod_i (2\pi\beta\tilde{J}_i)^{-1/2}\left[\sum_\sigma\int \beta d \phi _i \exp \left(-\beta H_i\right)\right] \equiv \prod_i Z_i ,
\end{equation}
and  $Z_D =\prod_i(\bar{J}_{ii}/\tilde{J}_i)^{1/2}$ comes from $H_D$.

The self-consistent renormalization up to $O(1/z_n)$ is carried out by requiring that the thermodynamic potential $\Omega$ of the system be stationary against independent variations of $\Pi _i$ and $a_i$.    
In accordance with the partition function we decompose $\Omega$ as
\begin{equation}
\Omega \{\Pi, a \} = \Omega_G +\Omega_L -\Omega_D ,
\end{equation}
where each suffix corresponds to that of the partition function.
We write $\Omega_L = \sum_i \Omega _i$ with
$$
\Omega_i = - T\ln  Z_i. 
$$
The simplicity of the Ising model allows us to obtain $\Omega_i $ analytically.
 Namely we first perform the $\phi$ integration and then sum over $\sigma$.  The result is given by
\begin{eqnarray}
\Omega _i & = & 
-\frac 12\tilde{J}_i (m_i^2+1)-T\ln [2\cosh \beta\bar{h}_i ],\\
\bar{h}_i & = & h_i +a_i-\tilde{J}_i m_i.
\label{eq:barh} 
\end{eqnarray}
We interpret $\bar{h}_i$ as the effective field at site $i$.  Note that $\tilde{J}_i m_i$ represents the reaction field \cite{brout} which is absent in the effective field of the MFA.
By variation with respect to $a_j$ we obtain
\begin{equation}
\frac{\partial \Omega}{\partial a_j} =
\sum_i [\delta _{ij}-\tilde{J}_i(J^{-1})_{ij}][m_i-\tanh (\beta\bar{h}_i)] = 0,
\label{eq:stationary_a}
\end{equation}
or equivalently
\begin{equation}
m_i = \tanh (\beta\bar{h}_i).
\label{eq:av of phi}
\end{equation}
We emphasize that the reaction field enters automatically in our theory as a consequence of the variational principle.  
On the other hand variation of $\Pi _i$ gives
\begin{equation}
\frac{\partial\Omega}{\partial\Pi _i}  =
\frac 12 [1-(\bar{J}_{ii} )^{-2}(\bar{J}^2)_{ii}]
[T\bar{J}_{ii}-\langle \delta\phi_i^2\rangle ] =0.
\label{eq:stationary_p}
\end{equation}
Here we have used the matrix notation for $\bar{J}^2$ and the average $\langle \delta\phi_i^2\rangle$ should be taken with reference to $Z_L$.
From eq.(\ref{eq:stationary_p}) we obtain
\begin{equation}
\langle\delta\phi _i^2\rangle =T\bar{J}_{ii}
= T\tilde{J}_i (1-\Pi_i\tilde{J}_i)^{-1} .
\label{eq:local average}
\end{equation}
Equation (\ref{eq:local average}) asserts that $\Pi _i$ gives the renormalization of the bare on-site interaction $\tilde{J}_i$ to $\bar{J}_{ii}$ just as it does the renormalization from $J_{ij}$ to $\bar{J}_{ij}$.

We now derive the susceptibility by using the formula:
\begin{equation}
\chi_{ij} +\beta m_im_j =\beta  \langle\sigma_i\sigma_j\rangle  
 = -2\beta\frac{\partial \Omega}{\partial J_{ij}} .
\end{equation}
In taking the derivative of $\Omega$ we care only the explicit dependence of $J_{ij}$ because of the stationary property represented by eqs.(\ref{eq:stationary_a}) and (\ref{eq:stationary_p}).
Then the contribution comes only from $\Omega _G$ giving
\begin{equation}
\chi_{ij} = [\Pi (1-J\Pi)^{-1}]_{ij},
\label{eq:chifluc}
\end{equation}
where the matrix notations are used.  
This result is consistent with the perturbation formula given by eq.(\ref{eq:dyson}).

We quote another relation which is obtained by partial integration of the quantity:
\begin{equation}
\sum_\sigma\int \scr D \phi \exp\left(-\frac \beta 2\sum (J^{-1})_{ij}\phi_i\phi_j\right)\frac{\partial ^2}{\partial\phi_i\partial\phi_j}\exp\left(
\beta\sum_i (\phi_i +h_i)\sigma_i \right).
\end{equation}
Namely we obtain
\begin{equation}
\langle \sigma_i\sigma_j\rangle 
= \sum_{lm}(J^{-1})_{il}\langle\phi_l\phi_m\rangle 
(J^{-1})_{mj}-T(J^{-1})_{ij}.
\label{eq:chi vs phi}
\end{equation}
This formula gives the rigorous relationship between the susceptibility and $\langle\phi_l\phi_m\rangle $.
Substituting  
$\langle\phi_l\phi_m\rangle =T\bar{J}_{lm} +a_la_m $ and $\bar{J}^{-1}= J^{-1}-\Pi $,
we obtain the same result for $\chi_{ij}$ as given by eq.(\ref{eq:chifluc}).
Thus we see that the single-site optimization is consistent with the general property given by eq.(\ref{eq:chi vs phi}).

By using eq.(\ref{eq:local average}) together with eq.(\ref{eq:chi vs phi}) applied to the effective single-site system, we obtain the local susceptibility $\chi_{ii}$ as
\begin{equation}
\chi_{ii} ^{-1} = \Pi_i ^{-1} -\tilde{J}_i.
\label{eq:local_vs_tilde}
\end{equation}
With matrix notations $\chi $ for the susceptibility, and $\chi _L$ for the local susceptibility, eq.(\ref{eq:chifluc}) is equally written 
with the help of eq.(\ref{eq:local_vs_tilde}) as
\begin{equation}
\chi ^{-1} =\chi _L^{-1} -J+\tilde{J},
\end{equation}
where $J$ and $\tilde{J}$ are also considered as matrices.
In the case of homogeneous magnetic field, the system acquires the translational invariance.
Then we should recover the local susceptibility by the momentum average of $\chi (\itv q) $.
The self-consistency relation is given by
\begin{equation}
1 =  \mathop{\rm Av}_{\itv q}\frac{1}{1-(J_{\itv q} -\tilde{J})\chi_L},
\label{eq:av=1}
\end{equation}
where ${\rm Av}_{\itv q} = N^{-1}\sum_{\itv q} $ and
 $J_{\itv q}$ is the Fourier transform of  $J_{ij}$.
One can also write the same relationship as
\begin{equation}
\frac{\tilde{J}}{1-\tilde{J}\Pi} = \mathop{\rm Av}_{\itv q} \frac{J_{\itv q}}{1-J_{\itv q}\Pi } .
\label{eq:local vs q-av}
\end{equation}

Now we mention a drawback of the SMA that the Ward-Takahashi identity (WTI) is violated. \cite{englert}
In order to see a consequence of the violation
we derive the susceptibility 
$ \chi _{ij} = \partial m_i/\partial h_j$ by direct differentiation. 
We obtain from eqs.(\ref{eq:barh}) and (\ref{eq:av of phi})
\begin{equation}
\chi _{ij} = \beta (1-m_i^2)\left[\delta _{ij} +\sum_l (J_{il}-\delta_{il}\tilde{J}_i)\chi_{lj}
-m_i\frac{\partial\tilde{J}_i}{\partial h_j}\right].
\label{eq:chi}
\end{equation}
The local susceptibility is given by $\chi _{ii} =\beta (1-m_i^2)$ independent of the interaction.  Hence   
we recover eq.(\ref{eq:chifluc}) only if we can neglect $\partial\tilde{J}_i/\partial h_j $. 
Fortunately this is indeed justified in the zero-field limit 
since the time reversal invariance requires $\tilde{J}$ to be an even function of magnetic field.
At finite field, however, the susceptibility depends on whether one derives it from fluctuation formula or from the thermodynamic derivative.
This drawback of the theory should be kept in mind if one discuss the property in finite magnetic fields. \cite{brout} 

The violation of the WTI originates from the site-diagonal property of the polarization function, and is very hard to remedy completely. 
The WTI requires consistency between quantities with different  orders of magnitude in $1/z_n$.
For example, all the site-diagonal self-energy diagrams in the fermion representation are included in the SMA since they are of $O(1)$ and $O(1/z_n)$.  
However, the WTI requires inclusion of polarization diagrams of $O(1/z_n^2)$ as well.   An example is the one shown in Fig.1(c) which is generated from an $O(1/z_n)$ part of the self-energy.
In the absence of magnetization, this diagram and related ones vanish by the particle-hole symmetry.  Thus one recovers the WTI in this limit.

\subsection{Entropy and specific heat}

The entropy $S$ of the system is derived by the standard formula $S = -\partial\Omega/\partial T$.
The stationary property of $\Omega$ leads to enormous simplification; one can neglect the implicit $T$-dependence of the variational parameters $a_i$ and $\Pi_i$.  Hence we obtain 
\begin{equation}
S = \frac 12 {\rm Tr} 
[\ln (1-\tilde{J}\Pi)-\ln (1-J\Pi)]+\sum_i  S_i,
\label{eq:c-entropy}
\end{equation}
where $S_i = \ln (2\cosh \beta\bar{h}_i)-\beta\bar{h}_im_i$ is the contribution of the spin at site $i$.
The form of $S_i$ becomes
the same as that of an isolated spin if one represents $S_i$ in terms of $m_i$.  Namely we have
\begin{equation}
S_i = \ln 2 -\frac 12 (1+m_i)\ln (1+m_i)-\frac 12 (1-m_i)\ln (1-m_i).
\label{eq:entropy}
\end{equation}
The specific heat $C = T(\partial S/\partial T)_h$ of the system is given by
\begin{equation}
C = -\frac T2 \sum_i \left[\chi _{ii}\frac{\partial\tilde{J}_i}{\partial T}
+\ln \left(\frac{1+m_i}{1-m_i}\right) \frac{\partial m_i}{\partial T}\right],
\label{eq:specific}
\end{equation}
where we have used the cancellation implied by eq.(\ref{eq:local vs q-av}). 

In the following we derive the specific heat explicitly in the case without magnetic field.  We set $m_i =0$ assuming the paramagnetic phase and remove the site indices of all local quantities.
At high temperatures, it is easy to derive the leading term as
$$ 
\tilde{J} = \mathop{\rm Av}_{\itv q} J_{\itv q} ^2 /T \equiv A/T.
$$
Then the entropy and specific heat is given by
\begin{equation}
S \sim \ln 2 - A/T^2, \ \ C \sim A/T^2.
\end{equation}
On the other hand, for general temperature it is convenient to represent $\partial\tilde{J}/\partial T $ in terms of susceptibilities.
The temperature dependence of $\tilde{J}$ is derived by the use of the self-consistency relation eq.(\ref{eq:av=1}).  
After some algebra we obtain a compact formula:
\begin{equation}
C = \frac N2 \left( 1-\frac{\chi _L^2}{\mathop{\rm Av}_{\itv q} \chi _{\itv  q}^2 } \right),
\end{equation}
which is assured to be positive by the inequality
${\rm Av}_{\itv  q} (\chi _{\itv  q}^2) \geq ({\rm Av}_{\itv  q}  \chi _{\itv  q} )^2 $.
It is evident from above that the specific heat and the susceptibility shows the anomaly at the same temperature,  and that $C$ in the paramagnetic phase becomes larger as intersite correlation develops.  This is in sharp contrast to the MFA where $C=0$ for $T>T_c$.
However, the critical property of the present theory is the same as the MFA.

\section{Generalization to Quantum Spins}

\subsection{Self-consistent equations for the Heisenberg model}

In this section we extend the formalism developed in \S 2 to quantum models.
As a representative of quantum models we consider the generalized Heisenberg model given by
\begin{equation}
H = -\frac 12\sum_{ij}J_{ij}{\itv S}_i\cdot {\itv S_j}
-\sum_{i}{\itv h}\cdot {\itv S}_i ,
\end{equation}
For clarity we assume in this section that the magnetic field is homogeneous.   
We emphasize that the formalism can be equally applied to systems without the translational invariance and with lower symmetry.
The path integral representation is accomplished by the coherent-state representation of spins.  
Equivalently one introduces fictitious fermions as 
\begin{equation}
\itv S _i = \sum_{\xa\xb}f_{i\alpha}^\dagger \itv\sigma _{\xa\xb} f_{i\xb},
\end{equation}
where $\itv \sigma$ is the vector composed of the Pauli matrices
and the constraint $\sum_\xa f_{i\alpha}^\dagger f_{i\alpha} =1$ is imposed. 
Note that $\itv S _i $ is twice the usual spin operator.   
Then the trace over spin degrees of freedom is accomplished by the path integral \cite{read} 
\begin{equation}
\int \scr D f^\dagger\scr D f \prod_i d\lambda_i \exp \left[- 
\sum_{i\xa}\int_0^\xb d\tau f_{i\alpha}^\dagger (\tau)\left( \frac{\partial}{\partial\tau}+i\zl _i\right)f_{i\alpha}(\tau) +i\zb\sum_i\zl_i 
\right] \equiv
\int \scr D\itv S \exp (-S_B),
\label{eq:spin-measure}
\end{equation}
where $f_{i\alpha}(\tau)$ and $f_{i\alpha}^\dagger(\tau)$ are the Grassmann numbers and $\zl _i$ is the Lagrange multiplier fields to enforce the constraint.
$S_B$ stands for the Berry phase term.

With this preliminary we obtain the expression of the partition function
\begin{equation}
Z  =  \int \scr D \itv S \exp [-S_B -\int_0^\xb d\tau H(\tau)].
\end{equation}
This form enables us to utilize the Gaussian identity eq.(\ref{eq:Gaussian}) for each imaginary time interval $\zD\tau$, since in expanding the exponential we can forget about the non-commuting nature of quantum operators.  As generalization from the classical case we introduce the time-dependent vector field $\itv\phi _i (\tau)$ for each site,  and write $Z$ as
\begin{equation}
Z  =  \int \scr D \itv S \scr D \itv\phi  \exp [-S_B -\int_0^\xb d\tau H_\phi(\tau)],
\end{equation}
where $H_\phi(\tau)$  is given by
\begin{equation}
H_\phi = \frac 12\sum_{ij} (J^{-1})_{ij}\itv\phi_i\cdot\itv\phi_j
-\sum_i (\itv \phi_i +\itv h)\cdot\itv S_i .
\end{equation}
Here all time-dependent quantities are specified at time $\tau$.  

The renormalization goes parallel to the classical case if we regard $J_{ij}$ as small quantities of $O(1/z_n)$.   Namely we introduce $\itv a$ as the static molecular field at each site.   The polarization function now has retardation effect, and  $\Pi (i\nu_n)$ represents a Fourier component where $\nu _n =2\pi n T$ is the Matsubara frequency with $n$ being an integer.  As in the Ising case, the polarization function is diagonal in site indices up to  $O(1/z_n)$.   
However, it has now two components $\Pi^\| (i\nu_n)$ and $\Pi^\perp (i\nu_n)$ in the presence of nonzero $\itv m$.
Accordingly the effective exchange interaction $\bar{J} _{ij} (i\nu_n)$ is given in the momentum space by
\begin{equation}
\bar{J}_{\itv q} (i\nu_n)^{-1} = J_{\itv q}^{-1} -\Pi (i\nu_n),
\end{equation}
which is understood as a $3\times 3$ matrix equation with the diagonal matrix $\Pi = {\rm diag}(\Pi^\perp,\Pi^\perp,\Pi^\|)$.  
Since $J_{\itv q}$ is a scalar in the Heisenberg model, we obtain the diagonal matrix $\bar{J}_{\itv q}$ as ${\rm diag}(\bar{J}_{\itv q}^\perp,\bar{J}_{\itv q}^\perp,\bar{J}_{\itv q}^\|)$. 

The thermodynamic potential is decomposed as $\Omega = \Omega_G +\Omega_L-\Omega_D $ in the same way as in the classical case.  
Writing $\bar{J}_D  (i\nu_n)\equiv \bar{J}_{ii} (i\nu_n)$ we obtain
\begin{eqnarray}
\Omega _G & = & -\frac T2\sum_{\itv q n} {\rm tr}\ln \left[\bar{J}_{\itv q}(i\nu _n)/J_{\itv q}\right] +\frac 12\sum_{ij}(J^{-1})_{ij}\itv a^2, \\
\Omega _D & = & -\frac T2 N\sum_{n} {\rm tr}\ln [\bar{J}_D(i\nu _n)\tilde{J}(i\nu _n)^{-1}],
\end{eqnarray}
where tr is the trace over the Cartesian indices, and $\tilde{J} (i\nu_n)^{-1} = \bar{J}_D (i\nu_n)^{-1} +\Pi (i\nu_n)$. 
The nontrivial part is $\Omega_L = -NT\ln Z_1 = N\Omega_1$ where the partition function $Z_1$ of the effective impurity is given by
\begin{equation}
Z_1 = \int \scr D \itv S_1 \scr D\itv \phi_1 \exp [-\int_0^\beta d\tau \scr L_1 (\tau )-S_{B1} ].
\end{equation}
Here $S_{B1}$ is the Berry phase term of the site and
\begin{equation}
\scr L_1 (\tau ) = \frac {1}{2\beta} \int_0^\beta d\tau '\delta\itv\phi (\tau) \cdot\tilde{J}(\tau -\tau ') ^{-1}\delta\itv\phi (\tau ') +
[\itv m -\itv S(\tau)]\cdot\delta\itv\phi (\tau) -(\itv h+\itv a)\cdot\itv S(\tau).
\label{eq:Lag1}
\end{equation}
In the above we have suppressed the obvious spatial index 1, and\begin{equation}
\tilde{J}(\tau ) = \sum_n \tilde{J} (i\nu_n)\exp( -i\nu_n\tau ).
\end{equation}
One has to solve the single-site problem explicitly to obtain the local susceptibility matrix $\chi_L(i\nu_n)$ for given $\tilde{J}(i\nu_n)$.  
By symmetry these quantities are diagonal with two independent components indexed by $\perp$ and $\|$.
In contrast to the Ising model, it is not possible to obtain $Z_1$ analytically.  
Instead, stationary conditions for $\Omega$ against variations of $\itv a$ and $\Pi (i\nu _n)$ lead to the following relations:
\begin{equation}
\int_0^\beta d\tau \langle \delta\itv\phi (\tau)\rangle =0, \ \ \  
\langle \delta\phi_\xa (\tau) \delta\phi_\xb (\tau ')\rangle = T\left[\bar{J}_D(\tau -\tau ')\right]_{\xa\xb}.
\end{equation}
Thus we obtain the set of equations which generalize those for the Ising model.  
For the two components $\zl = \perp ,\|$ we obtain
\begin{equation}
\chi _L^\xl (i\nu _n) = \Pi ^\xl (i\nu _n)[1-\tilde{J}^\xl (i\nu)\Pi ^\xl (i\nu _n)]^{-1} = \mathop{\rm Av}_{\itv q} \chi ^\xl (\itv q, i\nu _n).
\end{equation}
Here the $\itv q$-dependent dynamical susceptibility $\chi ^\xl (\itv q, i\nu _n)$ is given by
\begin{equation}
\chi ^\xl (\itv q, i\nu _n) = \Pi ^\xl (i\nu _n)[1-J_{\itv q}\Pi ^\xl (i\nu _n)]^{-1}
 = \chi ^\xl _L(i\nu _n)\left\{1-[J_{\itv q}-\tilde{J}^\xl (i\nu)]\chi^\xl _L (i\nu _n)\right\}^{-1} .
\end{equation}
Thus we obtain the self-consistency equation
\begin{equation}
1 = \mathop{\rm Av}_{\itv q} \left\{1-[J_{\itv q}-\tilde{J}^\xl (i\nu_n)]\chi_L ^\xl (i\nu _n)\right\}^{-1} ,
\label{eq:self-consistency}
\end{equation}
for each Matsubara frequency and each component $\zl$.
In \S 5 we generalize these equations to arbitrary localized configurations.

\subsection{Entropy}

In the quantum case the entropy $S$ of the system is conveniently derived by the formula 
$ TS = U-\Omega  $
where $U =\langle H\rangle $ is the internal energy of the system.  Another formula $S = -\partial \Omega/\partial T$ is less convenient in the quantum case because one also has to take the derivative of Matsubara frequencies.
Let us first express $\Omega_1$ by 
rearranging the perturbation terms as in
\S 2 or in the Fermi liquid theory. \cite{luttinger}  In the present case we regard the term $\delta\itv\phi\cdot\itv S$ as the perturbation.   
In order to perform the renormalized expansion we introduce
a fictitious matrix field $u(\tau)$ which increases  $\tilde{J}(\tau -\tau ') ^{-1}$ in eq.(\ref{eq:Lag1}) to $\tilde{J}(\tau -\tau ') ^{-1} +u(\tau -\tau ') $.  We obtain
\begin{equation}
2\delta \Omega_1 
= \sum_{\xa\xb} \langle 
\delta\phi_\xa (\tau) \delta\phi_\xb (\tau ') \rangle \delta u_{\xa\xb}(\tau -\tau ') 
= T\sum_{\xa\xb} \bar{J}_D (\tau -\tau ')_{\xa\xb}\delta u_{\xa\xb}(\tau -\tau '),
\label{eq:delta Omega_1}
\end{equation}
where $\bar{J}_D (\tau -\tau ')$ is generalized from its original meaning with $u =0$.  The Dyson equation represents the relationship in the general case as
$\bar{J}_D^{-1}  = \tilde{J}^{-1} + u-\Pi$, and  
we obtain
$\delta u = \delta\bar{J}_D ^{-1} +\delta \Pi$.  
Then with some manipulation similar to the one used in \S 2,  we can integrate eq.(\ref{eq:delta Omega_1}) from the decoupled limit $\bar{J}_D =\tilde{J}$ to the actual value of $\bar{J}_D$.  The result is 
\begin{equation}
2\Omega _1 \{\bar{J}_D\} = T\sum_n {\rm tr}\left\{-\ln [\bar{J}_D (i\nu_n)\tilde{J} (i\nu_n)^{-1}] +\bar{J}_D (i\nu_n)\tilde{J} (i\nu_n)^{-1} -1\right\}  
-T\Phi_1\{\bar{J}_D\} +2\tilde{\Omega}_0 ,
\label{eq:Omega1}
\end{equation}
where 
$$
\tilde{\Omega}_0 =  -T\ln (2\cosh \beta |\itv h +\itv a| ),
$$
is the MFA contribution.
The functional $\Phi_1\{\bar{J}_D\}$ satisfies the relation 
$ \delta \Phi_1/\delta \bar{J}_D(i\nu_n) = \Pi (i\nu_n) $.  
As in the classical case $\Phi_1\{\bar{J}_D\}$ is written symbolically as
\begin{equation}
\Phi_1\{\bar{J}_D\} = {\rm tr} \sum_{n=0}^{\infty} \frac{1}{n+1} \Pi_n \{\bar{J}_D\} \bar{J}_D,
\end{equation}
where $\Pi_n $ is a part of the polarization function made up of all $n$-th order skeleton diagrams with respect to $\bar{J}_D$, and the frequency summation is implicit.

We note that $\itv m$ and $\itv a$ are kept fixed during the variation, although actual change of the Hamiltonian does change the equilibrium magnetization.
This is because the procedure of variation is taken only to use the topological structure of perturbation processes. \cite{bloch}
We should also notice the stationary property $\delta\Omega _1/\delta\bar{J}_D =0$ at $\bar{J}_D^{-1}=\tilde{J}^{-1}-\Pi$ in eq.(\ref{eq:Omega1}).

The internal energy is given by
\begin{equation}
U = -\frac T2 {\rm tr}\sum_{\itv q n} J_{\itv q} \chi (\itv q, i\nu_n) 
-\frac 12 \itv m^2\sum_{ij}J_{ij}
-N\itv h\cdot\itv m .
\end{equation}
Then from $U-\Omega $ we obtain
\begin{equation}
S =  -\frac 12 \sum_{\itv q n}{\rm tr}\left\{
\ln [1-J_{\itv q}\Pi (i\nu _n)] +
 2\bar{J}_D(i\nu _n)\Pi (i\nu _n)\right\} + \frac N2 \Phi_1+ S_0,
\label{eq:q-entropy}
\end{equation}
where 
$ S_0 /N = -\beta (\itv h +\itv a)\cdot \itv m+\ln (2\cosh \beta  |\itv h +\itv a| )$.   
In contrast to the Ising case,
$S_0$ does not have the reaction-field correction $-\tilde{J}\itv m$.  The information of the reaction field is hidden in $\Phi_1$. 
Thus without solving the single-site problem explicitly, there is not much we can say generally.

Nevertheless the result eq.(\ref{eq:q-entropy}) is convenient to discuss the limiting behavior.  If the magnetic field is absent in the paramagnetic phase, we always have $S_0 = N\ln 2$.  
In the high temperature limit,  this term alone survives as it should.  
As temperature $T$ decreases,  the first term in eq.(\ref{eq:q-entropy})  makes positive contribution from extended spin fluctuations.  On the other hand the following terms up to $ N\Phi_1/2$ gives a negative contribution to correct the overcounting of on-site spin fluctuations.
If the entropy given by eq.(\ref{eq:q-entropy}) does not tend to vanish as $T$ goes to zero, the paramagnetic state becomes unstable against a magnetic order.  This is always the case for classical spins.
The possibility to have the paramagnetic ground state in the extreme quantum case is discussed in more detail later.

\subsection{Elimination of auxiliary fields}

Instead of performing the linked-cluster expansion as described above,  it is also possible to integrate away the $\phi$ fields in eq.(\ref{eq:Lag1}).  
The resultant Lagrangian $\scr L_s$ is given by (apart from the Berry phase term) 
\begin{equation}
\scr L_s(\tau) = 
-\frac {1}{2\beta} \int_0^\beta d\tau' \itv S (\tau) \cdot\tilde{J}(\tau -\tau ') \itv S (\tau ') 
-\frac 12 \itv m\cdot\tilde{J}(0)\itv m -
\itv S(\tau)\cdot (\itv h+\itv a -\tilde{J}(0)\itv m),
\label{eq:Lag_s}
\end{equation}
where the matrix $\tilde{J}(0)$ denotes the static component of $\tilde{J}(i\nu_n)$. 
Interestingly this Lagrangian contains explicitly the reaction field in contrast to the previous expansion.
In spite of its simple appearance, treatment of the Lagrangian is not easy because of the Berry-phase term. 
In the next section we show that the equivalent partition function is obtained by introducing a fictitious Hamiltonian.

\subsection{Static approximation}

If the temperature is much higher than $T_c$, only the static component $\nu_n=0$ of the Matsubara frequency remains important.  Then it is reasonable to neglect all the other components.
This is called the static approximation.
The static approximation in the present theory still keeps the non-commuting character of quantum spins.  In this respect it is different from our previous theory, \cite{uimin}  which replaced the spin vector composed of the Pauli matrices by a classical vector.

In the static approximation, the quantum spin sees a static external field which has the Gaussian distribution specified by $\tilde{J}$.  The path integral over $\itv S$ can be replaced by the trace of the density operator with
the quantization axis taken in the direction of $\itv h+\itv a+T\itv\xi$ with $\itv\xi =\beta \delta\itv\phi$.
We then have
\begin{equation}
Z_1 = \det (2\pi\beta\tilde{J})^{-1/2}\int d\itv\xi \exp \left( -\frac{T}{2}\itv\xi\cdot\tilde{J}^{-1}\itv\xi  -\itv m\cdot\itv\xi\right)
2\cosh \beta |\itv h+\itv a+T\itv\xi |.
\label{eq:Z_static}
\end{equation}
In contrast to the case of the Ising model,  integration over $\itv\xi$ does not lead to concise expression.
However, the result simplifies in some limiting cases.  First, by replacing the vector quantities by corresponding scalars,  eq.(\ref{eq:Z_static}) is reduced to
\begin{equation}
Z_1 = \exp[\frac \beta 2 \tilde{J}(1+m^2)] 2\cosh [\beta\bar{h}],
\end{equation}
where $\bar{h} = h +a-\tilde{J}m $ is the effective field with account of the reaction field.  Thus the result for the Ising model is reproduced.  

On the other hand, if one neglects $\partial/\partial\tau$ in the Berry phase term,  the non-commuting nature of spin operators is lost.  
Then we are left with spins behaving as classical vectors.
This approximation is justified in the limit of large spin $|\itv S|$.
If we further neglect the anisotropy in $\tilde{J}$,  we obtain from eq.(\ref{eq:Lag_s})
\begin{equation}
Z_1 =\int d\Omega_S \exp[\frac \beta 2 \tilde{J}(S^2+m^2) 
+\beta\itv S\cdot\bar{\itv h}],
\end{equation}
where $\bar{\itv h} = \itv h +\itv a-\tilde{J}\itv m $, and the integration is over the solid angle of $\itv S$.
Thus we obtain the partition function of classical dipoles with account of the reaction field.  The local susceptibility in this case is obtained as
\begin{equation}
\chi_L^\| = \beta S^2 [x^{-2} -\sinh ^{-2} x],
\hspace{1cm} 
\chi_L^\perp = \beta S^2 [x^{-1}\coth x -x^{-2}],
\end{equation}
where $\|$ ($\perp$) is the component parallel (perpendicular) to $\itv m$ and $x = \beta \bar{h}S$.

\section{Mapping to Spin-Boson Hamiltonian}

For numerical calculation, it is often convenient to work with a hypothetical Hamiltonian of the impurity spin interacting with bosons representing the effective medium.  
We shall show that the equivalent Hamiltonian is given by
\begin{equation}
H = \sum_{q\xl } \left[\omega_{ q\xl} b_{ q\xl}^\dagger b_{ q\xl} 
+ \frac{g}{\sqrt {N\omega_{ q\xl}}} (\itv m-\itv S)\cdot\itv e_{q\xl}\left( b_{ q\xl} +b_{q\xl}^\dagger\right)\right]
-(\itv h+\itv a)\cdot \itv S. 
\label{eq:spin-boson}
\end{equation}
This form is reminiscent of the spin-boson model investigated in another context. \cite{leggett,weiss,tsuzuki}
The same structure of perturbation diagrams is obtained whichever the boson field or the $\phi$ field is used.
Let the partition function of the impurity plus bosons be given by $Z_H$.
Then the partition function $Z_1$ of the effective impurity is given by 
 $ Z_1 = Z_H/Z_b$ where $Z_b$ is the partition function of bosons without the impurity.

In order to characterize the boson system leading to the equivalent partition function 
we replace $\delta\itv \phi (\tau)$ by boson operators as 
\begin{equation}
\delta \itv \phi (\tau) \rightarrow \sum_{q\xl} \frac{g}{\sqrt {N\omega_{q\xl} }}\itv e_{q\xl}\left[ b_{ q\xl}(\tau)+b_{q\xl}^\dagger (\tau)\right],
\end{equation}
where $b_{ q\xl}(\tau)$ is the annihilation operator of a boson with polarization $\itv e_{q\xl}$, and $g$ is the coupling constant.  The role of $g$ is only to adjust the dimension, and hence can be taken unity in practical calculation.  
Without loss of generality we can take the one-dimensional boson system with $q >0$.
The correlation function is replaced by the bare boson Green function $D_\xl ( q, i\nu _n)$.    
\begin{equation}
\langle \delta\phi _\xa(\tau) \delta\phi_\xb (0)\rangle 
\rightarrow 
 \frac{g^2T}{N} \sum_{ q\xl}\sum_n \frac{e_{q\xl}^\xa  e_{q\xl}^\xb}
 {\omega _{ q\xl}}D_\xl ( q, i\nu _n)\exp (-i\nu_n\tau),
\end{equation}
where 
$ D_\xl ( q,i\nu_n) = (i\nu_n -\omega _{ q\xl})^{-1}-(i\nu_n +\omega _{ q\xl})^{-1} $
with $\omega _{ q\xl}$ being the boson frequency.   
We write $\zl $ as $\perp$ (two-fold degenerate) and $\|$ according to the anisotropy introduced by the finite magnetization.
Then the bare spectral intensity is given by
\begin{equation}
\frac 1\pi {\rm Im}\tilde{J}^{\xl}(\omega) =  \sum_{ q} \frac{g^2}{N\omega _{ q\xl}}[\delta (\omega -\omega _{ q\xl})-\delta (\omega +\omega _{ q\xl})] = 
\frac{g^2 a_L}{2\pi\omega_{ q\xl} }\left(\frac{dq}{d\omega_{ q\xl}}\right),
\label{eq:JvsB}
\end{equation}
with $q$ satisfying $\omega_{q\xl} =\omega$ and $a_L$ the lattice constant in the rightmost expression.  
Here we have assumed that $\omega_{q\xl}$ increases monotonically with $q$.  
Thus for given ${\rm Im}\tilde{J}(\omega)$ and $g$, the spectrum $\omega_{q\xl}$ is derived from eq.(\ref{eq:JvsB}).
For example, if the boson has a spectrum  $\omega_{q\xl} \propto q^{1/p}$ for small $q$, we obtain
$  {\rm Im}\tilde{J}^{\xl}(\omega) \propto \omega^{p-2}$ for small $\omega$.

The ground state of the impurity without magnetic field becomes either doublet or singlet depending on the details of $\tilde{J} (\tau)$.
In the former case, the self-consistent solution actually drives the ground state to magnetic ordering.  
On the contrary, if the effective impurity has the singlet ground state,  the system as a whole also has the singlet ground state.  
This singlet state is reminiscent of the RVB picture. \cite{andersonR}   
Technically eq.(\ref{eq:spin-boson}) is not exactly the same as the standard spin-boson model which is anisotropic in the spin space.
In the bosonization approach to the Kondo problem, \cite{schotte,schotte2} one ends up with an equivalent classical two-component gas in one dimension with the interaction decaying as inverse square of the distance.
In our case,  
the long-time behavior is given by $\tilde{J} (\tau) \propto  \tau ^{-2}$ if ${\rm Im}\tilde{J}(\omega ) \propto \omega $ for small frequencies.  However, the resultant model is not an Ising model as seen from eq.(\ref{eq:Lag_s}).
It is possible to convert the model to an Ising-like one by 
eliminating the interaction part with $S_z$ by a canonical transformation. \cite{schotte}
Then $\partial S_z /\partial \tau$ plays the role of instantons.
 It remains to see how the self-consistency condition determines the actual shape of $\tilde{J} (\tau)$.

Among appropriate methods to solve the effective impurity problem,  we mention numerical methods such as the numerical renormalization group,\cite{sakai} the quantum Monte Carlo \cite{georges,jarrel}, or the resolvent method\cite{kim} all of which respect the strong on-site correlation.  
It seems that the most convenient way to obtain the self-consistent solution is to use the numerical iteration. 
The iterative procedure starts from a trial Im$\tilde{J}(\omega )$ and the resultant $\tilde{J}(i\nu_n)$. Then explicit solution of the Hamiltonian gives $\chi_L(i\nu_n)$, and hence $\Pi (i\nu_n)$.  Substituting these single-site quantities to the self-consistency equation (\ref{eq:self-consistency})
gives the second trial $\tilde{J}(i\nu_n)$ by
\begin{equation}
\tilde{J}(i\nu_n) = \Pi (i\nu_n)^{-1} -\left\{
\mathop{\rm Av}_{\itv q} [\Pi (i\nu_n)^{-1} -J_{\itv q}]^{-1}\right\}^{-1}.  
\end{equation}
Then one continues the second step of the iteration.

\section{Application to Higher Multipoles} 

The present formalism is flexible enough to be applied to various systems of localized electrons with two-body interactions.  
Let a local electronic configuration $a$ at site $i$ be represented by $|i,a\rangle$.  Then the transition to this state from
another configuration $|i,b\rangle$ is described by the Hubbard operator
$ X_i^{ab} = |i,a\rangle\langle i,b|. $  In the case of $a =b$ the $X$-operator is equivalent to the projection operator onto the state $a$.  For notational simplicity, we 
work with hermitian operators $X_i^{\{ab\}}\equiv (X_i^{ab}+X_i^{ba})/2, X_i^{[ab]}\equiv (X_i^{ab}-X_i^{ba})/(2i)$ and write them as  $X_i^{\xa}$ where $\alpha$ represents either $\{ab\}$ or $[ab]$.
Then we consider the Hamiltonian given by
\begin{equation}
H = -\frac 12\sum_{ij}\sum_{\xa\xb} J_{ij}^{\xa\xb}X_i^\xa X_j^\xb 
+\sum_{i a}\ze _a X_i^{aa},
\end{equation}
where the second term represents energy levels with possible splittings caused by magnetic field and/or by crystalline electric field (CEF).
The exchange interaction has a symmetry much lower than the point-group symmetry at each site.  Thus it has many nonzero elements with off-diagonal indices in general.  

For the partition function we use the path integral over $X_i^\xa$ which can again be performed with use of fictitious fermion or boson operators.  As in the spin case we impose the constraints that the sum of the projection operators be unity at each site.
The fluctuating field $\phi_i^\xa$ has the same set of indices as the $X$-operators. 
Then the partition function is given by
\begin{equation}
Z  =  \int \scr D X\scr D \phi \exp \left[-S_B -\int_0^\beta d\tau H_{\phi} (\tau)\right],
\end{equation}
where $S_B$ is the Berry-phase term and
\begin{equation}
H_\phi = \frac 12\sum_{ij}\sum_{\xa\xb} 
(J^{-1})_{ij}^{\alpha\beta}\phi_i^\alpha\phi_j^\beta 
+\sum_{i\alpha} \phi_i^\alpha X_i^\alpha 
+\sum_{i a}\ze _a X_i^{aa}.  
\end{equation}
Here $J^{-1}$ is the inverse of the double-indexed matrix composed of $J_{ij}^{\xa\xb}.$
One can follow the same procedure as in \S 4 to renormalize the Hamiltonian up to $O(1/z_n)$.  

We define the Green function of the fluctuating field by
$$
T\bar{J}_{ij}^{\alpha\beta}(\tau -\tau ')=\langle \phi_i^\alpha(\tau)\phi_j^\beta (\tau ')\rangle _c  \equiv \langle \phi_i^\alpha(\tau)\phi_j^\beta (\tau ')\rangle -\langle \phi_i^\alpha(\tau)\rangle\langle\phi_j^\beta (\tau ')\rangle ,
$$
where the average is to be taken over the distribution specified self-consistently.  To derive this we introduce the polarization matrix $\Pi (i\nu_n)$ by 
$ \bar{J}^{-1} = J^{-1} -\Pi $
where all quantities are matrices.
The important point to note is that $\Pi^\alpha (\tau)$ is diagonal both in site and $\alpha$.  The latter follows from the point group symmetry.  
By the same reason the matrix $\bar{J}_D$ is also diagonal with respect to the index $\alpha$.

The renormalization is carried out by requiring that matrix of the renormalized single-site propagator in $Z_1$ be equal to $\bar{J}_D$.  The self-consistency equation is given by
\begin{equation}
1 = \frac 1N\sum_{\itv q} \left\{ 1-[J_{\itv q} -\tilde{J}(i\nu_n)]\chi_L(i\nu_n)\right\}^{-1},
\label{eq:sc_CEF}
\end{equation}
where 1 in the left hand side is the unit matrix.
The matrix $J_{\itv q}$ is composed of the Fourier transform of  $\{J_{ij}^{\xa\xb}\}$, and $\chi_L(i\nu_n)$ is the local dynamical susceptibility matrix.
The effective interaction matrix $\tilde{J}(i\nu_n)$ satisfies the relation
$
\chi_L^{-1} = \Pi^{-1} -\tilde{J}.
$
The full dynamical susceptibility matrix $\chi (\itv q, i\nu_n)$ is given by
\begin{equation}
\chi (\itv q, i\nu_n) = \Pi (i\nu_n)[1-J_{\itv q}\Pi(i\nu_n)]^{-1}.
\end{equation}
These relations are just generalization of those derived in previous sections.  

The matrix equation is greatly simplified if one uses the basis set of irreducible representations.  
We sketch the procedure assuming the spherical symmetry around the effective impurity.  This is merely to make notations familiar,  and can be straightforwardly generalized to the point-group symmetry. \cite{kuramoto83}
The irreducible tensor operators $X(lm)$ are given by
\begin{equation}
 X(lm) = \sum_{ab} (2J+1)^{-1/2} \langle Ja|lm Jb\rangle X^{ab},
\end{equation}
where $\langle Ja|lm Jb\rangle $ is the Clebsch-Gordan coefficient.  Then the local susceptibility matrix in this basis set has only diagonal elements as given by
\begin{equation}
\langle lm|\chi _L(i\nu_n)|l'm'\rangle = \delta_{ll'}\delta_{mm'}\chi_L^{(l)}(i\nu_n).
\end{equation}
Since $l$ runs from 0 to $2J$, we have $(2J+1)$ independent components for $\chi _L(i\nu_n)$.  The same number of components are present also for $\Pi(i\nu_n)$ and $\tilde{J}(i\nu_n)$.
Thus eq.(\ref{eq:sc_CEF}) is reduced to $(2J+1)$ scalar equations.

It is convenient to introduce the Hamiltonian of an impurity coupled with fictitious bosons as in the case of the Heisenberg model.
The Hamiltonian without symmetry breaking fields is given by
\begin{equation}
H = \sum_{q\xl } \left[\omega_{ q\xl} b_{q\xl}^\dagger b_{ q\xl} 
+ \frac{g}{\sqrt {N\omega_{ q\xl}}} 
X_{\xl}\left( b_{ q\xl} +b_{\xl}^\dagger\right)\right],
\end{equation}
where the mode $\zl$ represents the set $(lm)$ of the irreducible tensor.
We note that the phase transition to the ordered phase is signaled by
\begin{equation}
\det [1-J_{\itv q}\Pi(0)] =0,
\label{eq:det}
\end{equation}
which reduces to the MFA if one uses the zero-th order result for the polarization function.  Namely all the fluctuation effect is encoded in $\Pi(0)$ in the present theory.
As temperature decreases,  the determinant becomes zero first for the most favorable order.   If $J_{\itv q}$ can be regarded as a scalar, there is no coupling among different irreducible representations.
The opposite extreme case is that $\itv q$ is situated at a low symmetry point of the Brillouin zone.  Then many components of $\Pi(0)$ couple in eq.(\ref{eq:det}).

\section{Discussion}

We compare the present theory with previous effective medium theories for fermions.  Both theories have similar sets of self-consistent equations.   Similarity to the fermion theory is seen by the correspondence $ \bar{J} \Leftrightarrow G$ and $\Pi \Leftrightarrow \Sigma $
where $G$ is the single-particle Green function and $\Sigma$ the self-energy of fermions.
The intersite exchange interaction in fermion models is generated by virtual particle-hole pair excitations. Hence 
a loop of exchange interactions, which is taken into account here in deriving the polarization function,  is equivalent to a loop of particle-hole bubbles in the fermionic theory.  
However, the intersite processes so far included in the fermionic theory just corresponds to tree diagrams of particle-hole bubbles.
Thus for magnetic properties the fermionic theory is still in the mean-field level.  
In order to study the itinerant magnetism 
beyond the mean-field theory,  it seems more convenient to use 
the {\it t-J} model instead of the Hubbard model.  
Then the single-site optimization of not only the polarization function but the fermion self-energy is accomplished by the use of Grassmann auxiliary fields in addition to the $\phi$ fields. 
 This subject will be treated in a separate paper. 

Concerning the possible Kondo-type effect, 
the Heisenberg model with only the nearest-neighbor interaction $J$ in the unfrustrated lattice is not a good candidate.   
As compared to the characteristic scale $z_n J$ for the Neel state,  the singlet cannot take advantage of many neighbors and the characteristic energy is $J$.
Hence the ground state of the nearest-neighbor model becomes the Neel state in high dimensions.
On the contrary, if there is substantial frustration in the longer-ranged exchange interaction, the critical temperature for the magnetic ordering may be suppressed severely.
The Kondo temperature is determined by the easiness to form singlet pairs between many neighbors.
Hence presence of the frustration should work differently.  
As the simplest example leading to the singlet ground state we consider the case where $J_{ij}$ is a negative constant $-K$ independent of the distance.
Then the Heisenberg model can be solved exactly since the energy is given by
\begin{equation}
E = 2K \itv S^2 -C,
\end{equation}
where $\itv S =\sum_i\itv S_i/2$ is the total spin of the system and $C=(3/2)KN$.  Thus any singlet state gives $E=-C$ as the ground state energy.  
Slight modification of the range of exchange interaction lifts the degeneracy among various singlets.  It appears possible that certain modification keeps one of the singlet states as the ground state.

Another important parameter to control the competition between different ground states is the number of components.  This number $n$ increases from 1 for the Ising model to 3 for the Heisenberg model.  The number 3 is interpreted as $2^2-1$ where 2 is the spin degeneracy and $-1$ is to remove the identity operator.
In the case of four-fold degenerate CEF states as in CeB$_6$, we have $n =4^2 -1=15$. \cite{shiina}
 In the case of TmTe, on the other hand, the CEF splitting seems to be very small as judged from the observed Curie constant\cite{matsumura} which is very close to that of Tm$^{2+}$, 
 and the low-energy features observed in the inelastic neutron scattering. \cite{mignot} 
If we assume the degeneracy of the Hund-rule ground state with $J=7/2$, the number of components is as large as $n=8^2-1 = 63$.
Although the susceptibility matrix breaks up into its irreducible components, 
the overall number of components is still a crucial parameter because of the constraint $\sum_a X_{aa} =1$.
In a forthcoming paper we shall report on numerical results of the static approximation for the quadrupole order. \cite{fukushima}
It is found there that the transition temperature to the quadrupole order is more strongly reduced from the MFA value as the number of components increases.

In summary, we have presented a self-consistent dynamical theory for quantum spins and multipoles.  The theory is a natural generalization of the spherical model approximation for classical models.  
Detailed dynamical results obtained by numerial calculation will be presented in future papers.

\section*{Acknowledgments}

The authors thank T. Tsuzuki, O. Sakai and H. Kusunose for helpful conversations on the effective single-impurity model. 
One of the authors (Y.K) is grateful to B. M\"{u}hlschlegel and J. Zittartz for information on the spherical model.  
Thanks are also due to E. M\"{u}ller-Hartmann for discussing about fermion theories in large dimensions.

\end{document}